\documentstyle[prl,aps,epsfig]{revtex}

\begin{document}

\twocolumn[\hsize\textwidth\columnwidth\hsize\csname @twocolumnfalse\endcsname

\title{The Shapiro Conjecture: Prompt or Delayed Collapse in the head-on 
collision of neutron stars?}

\author{Mark Miller${}^{(1)}$, Wai-Mo Suen${}^{(1,2)}$ and Malcolm
Tobias${}^{(1)}$}

\address{${}^{(1)}$McDonnell Center for the Space Sciences,
Department of Physics,
Washington University, St. Louis, Missouri 63130}

\address{${}^{(2)}$Physics Department,
Chinese University of Hong Kong,
Hong Kong}

\date{\today}
\maketitle

\begin{abstract}

We study the question of prompt vs. delayed collapse in the head-on
collision of two neutron stars.  We show that the prompt formation of a
black hole is possible, contrary to a conjecture of Shapiro which
claims that collapse is delayed until after neutrino cooling.  We
discuss the insight provided by Shapiro's conjecture and its
limitation.  An understanding of the limitation of the conjecture is
provided in terms of the many time scales involved in the problem.
General relativistic simulations in the Einstein theory with the full
set of Einstein equations coupled to the general relativistic
hydrodynamic equations are carried out in our study.

\end{abstract}


\narrowtext

\vskip2pc]

\paragraph*{\bf Introduction.}
\label{introduction}

The study of the coalescence of neutron stars (NSs) is important for
gravitational wave astronomy and high energy astronomy.  However, at
present we lack even a qualitative understanding of the process.  One
issue is the prompt vs. delayed collapse problem.  While we expect
that two $1.4$ $M_{\odot}$ NSs when merged will eventually collapse to
form a black hole, the collapse could be delayed by fragmentation/mass
shredding, angular momentum hang-up, and/or shock heating.  The time
scale of the collapse has important implications on the gravitational
wave signals to be detected by LIGO \cite{LIGO3}.  We focus on the
issue of prompt vs. delayed collapse in this paper.

The difficulty of getting an answer to this aspect of NS
coalescence is very much the same as the full coalescence 
problem.  Namely, we need to
solve the full Einstein equations coupled to the general relativistic
hydrodynamic (GR-hydro) equations.

Recently, Shapiro \cite{Shapiro98a} put up an argument suggesting that
one may be able to answer this question without numerical simulations,
at least for the case of head-on collisions.  The ``Shapiro
conjecture'' goes as follows: Given the conditions: (I) that the two NSs
are colliding head-on after falling in from infinity, and (II) the NSs
are described by a polytropic equation of state (EOS) $P = K \rho ^ \Gamma$
(with $K$ a function of the entropy and the polytropic index $\Gamma$
remaining constant throughout the collision process), it is conjectured
that no prompt collapse can occur for an arbitrary $\Gamma$ and an
arbitrary initial $K$.  The basic argument is that the potential
energy when converted to thermal energy by shock heating is always
enough to support the merged object, until neutrino cooling sets in.

The argument based on conservation is appealing, and provides useful
understanding for a range of the NS coalescence problems.  However
there is a major assumption for the argument to go through, namely,
the collision process can be approximated by a quasi-equilibrium
process, in two senses: (A) The coalescing matter can be described by
one {\it single} EOS everywhere ($K$ is a function of time but not
space), and (B) whether it collapses or not is determined by
hydro-static equilibrium conditions, i.e., whether a stable
equilibrium configuration {\it exists} or not.  This quasi-equilibrium
assumption is not self-evident for the head-on collision of heavier
NSs.  It could happen that the coalesced object collapses before it
can thermalize, {\it or} the collision process is so dynamic that even
though a stable equilibrium state exists, it is not attained in the
collapse process.  The final outcome depends on the various time
scales in the problem.

\paragraph*{\bf Time Scale Considerations.}
\label{time}
We examine this assumption of ``quasi-equilibrium''
and see if it can be justified under the conditions of (I) and (II) above.
We note that the collision process involves many time scales, and
there are at least six of them relevant for our present consideration:
1. The time scale associated with the infall velocity: $t_i = R/ V_i$, R=the
radius of the NS,  $V_i$= (infall velocity at the point of contact).
2. The time scale associated with the local sound velocity: $t_s =
R/ V_s$ ,  $V_s$= (sound velocity).
3. The time scale associated with the velocity of the shock (velocity in the
rest frame of fluid): $t_{sh} = R/V_{sh}$,  $V_{sh}$=(shock velocity).
4. The time scale for the merged object to 
thermalize, in the sense of being describable by one single EOS (same
$K$ everywhere): $t_e$.
5. The time scale of neutrino cooling $t_{n}$.
6. The time scale of the gravitational collapse $t_{c}$.

Some comments of these time scales are in order.  
We focus on the case of two 1.4
$M_{\odot}$ NSs.  We model them with a polytropic EOS
$P=K \rho^\Gamma$ with a polytropic index of $\Gamma
=2$.  The initial  
$K$ value of the two stars is taken to be 
$1.16 \times 10^5 \; \frac {{cm}^5}{g \; s^2}$.  (Maximum stable mass
of these values of $K$ and $\Gamma$ is 1.46 $M_{\odot}$.)
We note that the argument in \cite{Shapiro98a} is applicable to 
all polytropic models.

For this model, $V_i$ is (somewhat larger than) the Newtonian value
$\sim 0.28c$, as can be estimated by $\sqrt {GM/(2R)} $; the diameter
of the NSs is about $2R=26 km$ (the isotropic coordinate radius of
this NS is $9.3 km$, the proper radius is $R=13 km$).  Hence the time
scale associated with the infall velocity $t_i$ is about (smaller
than) $ 0.16 \; ms$.  To estimate the second time scale $t_s$, note
that the sound velocity $V_s$ depends strongly on the dynamical
process and the region under consideration.  For the model mentioned
above, the initial central rest mass density of the NSs is about $1.5
\times 10^{15} g/cm^3$; $V_s$ there is about 0.5 $c$.  With the
density elsewhere initially lower than this value, but higher in some
period in the central region of the collision, $V_s$ varies but is
roughly $0.5 c$.  Thus, $t_s$ is roughly $0.1 ms$.  To estimate the
third time scale $t_{sh}$ requires an estimation of the velocity of
the shock $V_{sh}$ produced in the collision.  The locally measured
proper velocity of the shock $V_{sh}$ is higher than, but of the same
order of magnitude of, the sound speed $V_s$ at a fraction of $c$ in
the head-on collision case.  Hence $t_{sh}$ is also of order $0.1ms$.
These three time scales determine the time scale 4 which is central to
our discussion.  In near static situation, or when the bulk velocity
of matter is small ($V_i << V_s$ and $V_i << V_{sh}$), $t_e$ can be
taken to be a few times $t_s$ or $t_{sh}$.  On the other hand, its
value in a highly dynamic situation with $V_i$ comparable to $ V_s$
and $V_{sh}$ is an important issue to be discussed below.  The fifth
time scale $t_n$ governs the final settling down of the merged object
after $t_e$.  $t_n$ is of the order of seconds, orders of magnitude
longer than the first four time scales.  The gravitational collapse
time scale $t_c$ is in turn controlled by these time scales 1-5.  It
can be as short as $t_i$, or as long as $t_n$.  For the collision of
two 1.4 $M_{\odot}$ NSs, the merged object would have to collapse
after $t_c$, if not before, for most of the reasonable EOS.  We call
collapse that occurs on the first four time scales prompt collapse,
and collapse that occurs on a longer time scale, like $t_n$, delayed
collapse.  For more general coalescence processes, there can be other
time scales involved, e.g., the time scale of angular momentum
transfer $t_a$, and the time scale of gravitational wave emission
$t_g$.  However, for the case of head-on collision with the stars
falling in from infinity, we expect strong shock heating causing $t_n$
to be shorter than $t_g$.  We do not have to consider $t_a$ and $t_g$
in our present consideration.  

In Shapiro's argument, the time scale 4, $t_e$, is implicitly taken to
be the shortest time scale in the problem, so that the system can be
described by a single EOS at any instant in the collision process.  The
above discussion suggests that this may not be true for the two 1.4
$M_{\odot}$ NS collision case.  Indeed, the relations between the time
scales 1, 2 and 3 strongly affect $t_e$.  With $t_i$ comparable to
$t_s$ and $t_{sh}$, dynamic effects are important, and $t_e$ can be
longer than $t_i$.  In particular, with matter falling
in at high speed along the axis of the collision, the speed of the
shock wave in that direction would be significantly reduced, until
{\it after} $t_i$, delaying ``thermalization'' of the coalescing
objects.  For situations like this, arguments based on a uniform EOS
throughout the coalescing object cannot be justified.  Indeed, when the
infalling time scale $t_i$ is comparable to the other time scales in
the process, it could happen that even if a hydrostatic stable
equilibrium configuration exist, the dynamics of the system might not
lead to that configuration and the time scale of collapse could be as
short as $t_i$.

Another way of looking at the problem is to imagine we tie the two stars on
strings and lower them towards one another in a quasi-stationary
fashion while depositing the potential energy extracted back to the
two stars.  For this case Shapiro's argument would be applicable.
However, for a NS collision with the time scales discussed
above, one would have to examine the dynamics of the infall to
determine whether a prompt or delayed collapse would occur.  In
short, as both a thermally supported merged object and a black
hole can have the same rest mass and total energy, arguments based
solely on conservation of mass and energy without taking the dynamics
into consideration cannot rule out one outcome from the other.

We note that the above time scale considerations suggests that whether it is
a delayed or prompt collapse in head-on collision can depend on
the initial NS's configuration.  It does {\it not} imply 
prompt collapse by itself.  To demonstrate that
a prompt collapse results, one has to perform a fully relativistic
simulation.

Our NCSA/Potsdam/Wash U collaboration is developing a multi-purpose 3D
numerical code, ``Cactus'', for relativistic astrophysics and
gravitational wave astronomy.  This code contains the Einstein
equations coupled to the general relativistic hydrodynamic equations.
For a description of various aspects of the code and the NS grand
challenge project based on it see \cite{Potsdam-WashU-web}.  Testbeds and
methods for evolving neutron stars have been given in \cite{Font98b},
and will not be repeated in this paper.  While this multi-purpose code
is still under development for various capabilities in treating a
broad class of astrophysical scenarios, in this paper we focus on the
results obtained by applying this code to the head-on collision
problem.

\paragraph*{\bf Simulation results.}
\label{results}

We show the $M= 1.4$ $M_{\odot}$ head-on collision case.  The
stars are modeled as given above.  We put the two TOV solutions at a
proper distance of $d=44 \; km$ apart (slightly more than 3 $R$
separation) along the z-axis, and boost them towards one another at
the speed (as measured at infinity) of $\sqrt {GM/d}$ (the Newtonian
infall velocity).  The metric and extrinsic curvature
of the two boosted TOV solutions are superimposed by 
(i) adding the off-diagonal components of the
metric, (ii) adding the diagonal components of the metric and
subtracting 1, and (iii) adding the components of the extrinsic
curvature.  The resulting matter distribution, momentum distributions,
conformal part of the metric, and transverse traceless part of the
extrinsic curvature are then used as input to York's procedure
\cite{York79} for determining the initial data, in maximal slicing.
With this setup, the initial data satisfies the complete set of
Hamiltonian and momentum constraints to high accuracy (terms in the
constraints cancel to $10 ^ {-6}$),
{\it and} physically represent two NSs in
head-on collision falling in from infinity, at least up to the
Newtonian order.  (For initial data setup on the P1N level, see
\cite{Shinkai99a}).  

The initial data is then evolved with the numerical methods described
in \cite{Font98b}.  Various singularity avoiding slicings been used
and tested against one another (maximal and $1+log$ slicings most
extensively), yielding basically the same results.  The simulations
have been carried out with resolutions ranging from $\Delta x
= 1.48 \; km$ to $0.246 \; km$ (13 to 76 grid points across each NS,
with $32^3$ to $192^3$) for convergence and accuracy analysis.

% Figure
\begin{figure}
\psfig{figure=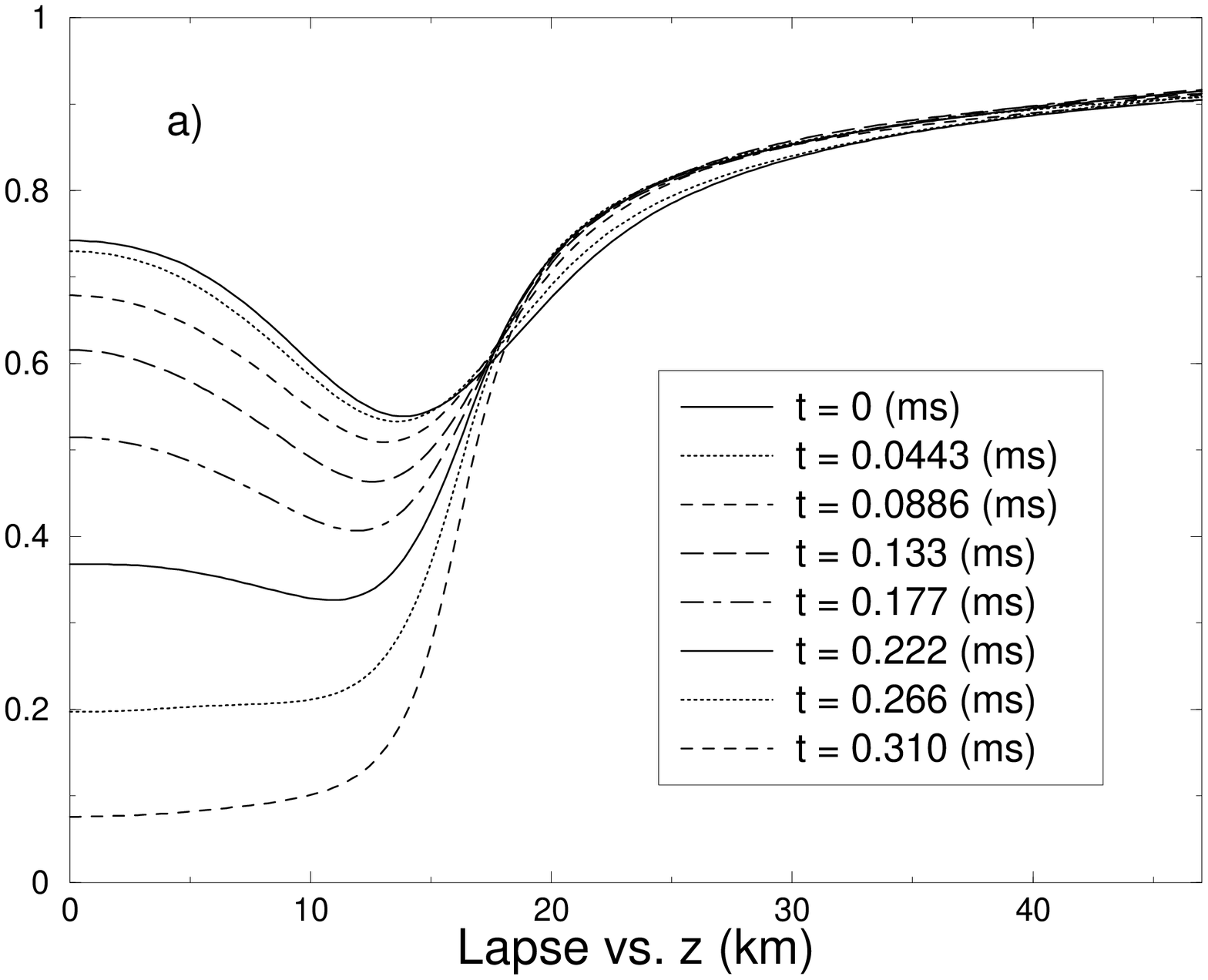,width=4.0cm}
%\vspace{-3.2cm}
%\hspace{-0.1cm}
\psfig{figure=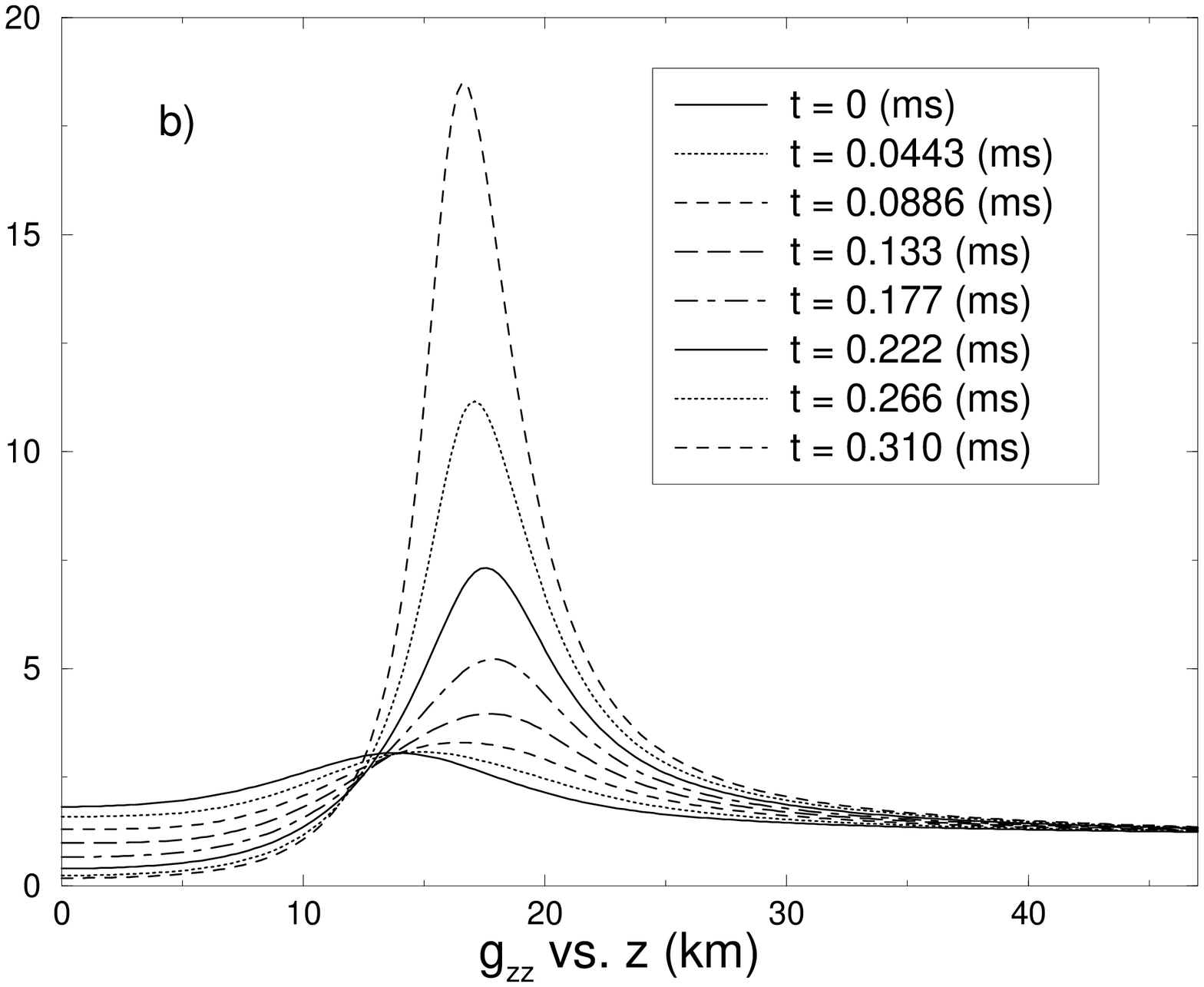,width=4.0cm}
\caption{
   a) The lapse ($\alpha$) and b) $g_{zz}$ along the $z$-axis 
      are displayed at
      various times.  This simulation used
      $192^3$ grid points, with $\Delta x = 0.246 \; km$.  
}
%\label{fig:alpgzz}
\end{figure}

In Fig. 1a we show the collapse of the lapse along the
$x=y=0$ line from $t=0 \; ms$ to $t=0.31 \; ms$ at
intervals of $0.044 \; ms$. (With the reflection symmetry across the $z=0$ plane
and the axisymmetry of the head-on collision, we only need to
evolve the first octant.)  At
$t=0.31 \; ms$ the lapse has collapsed significantly.

In Fig. 1b we show
the evolution of $g_{zz}$ along the $z$ direction ($x=y=0$).  
We see the familiar
grid stretching effect associated with evolving a black hole.

% Figure
\begin{figure}
\psfig{figure=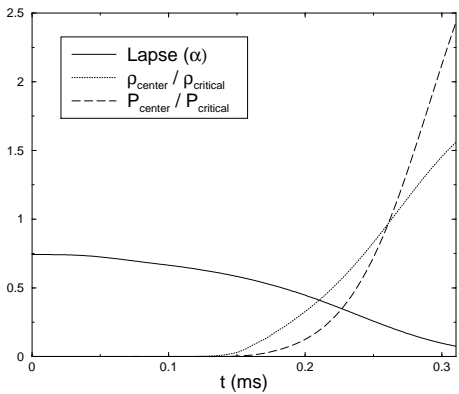,width=6.0cm}
\caption{ The evolution of the lapse ($\alpha$), the rest mass density
          ($\rho$), and the pressure ($P$) in the region centered
          at the point $x=y=z=0$ to the time 
          $t = 0.31 \; ms$.  
}
\label{fig:lapserhop}
\end{figure}

Fig. 2 shows the time development of the lapse, the (proper) rest mass density
$\rho$ and the pressure $P$ at the origin, scaled by the critical
secular stability values $\rho_{critical}$ and $P_{critical}$, the
values beyond which a static TOV solution is unstable to collapse for
the given polytropic coefficient $K$ and index $\Gamma(=2)$.  We note
that the effective $K(=P/ \rho ^2)$ is time dependent due to shock heating.
At coordinate time $t=0.26 \; ms$ we 
see that both $\rho$ and $P$ surpass $\rho_{critical}$ and
$P_{critical}$, indicating a collapse.

% Figure
\begin{figure}
%\hspace{5.0cm}
\psfig{figure=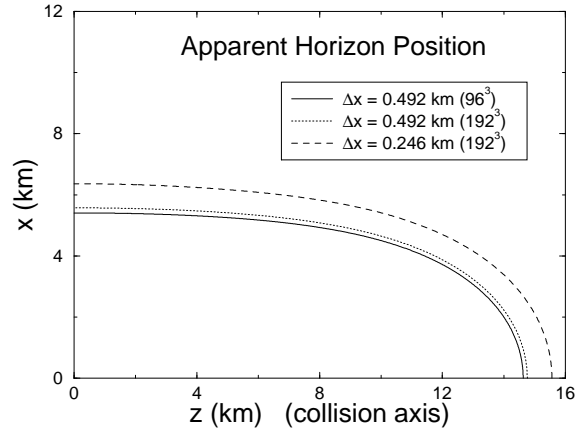,width=7.5cm}
\caption{ The position of the AH at different resolutions and outer
boundary locations, all at $t = 0.31 \; ms$.
}
\label{fig:horizon}
\end{figure}

% Figure
\begin{figure}
\psfig{figure=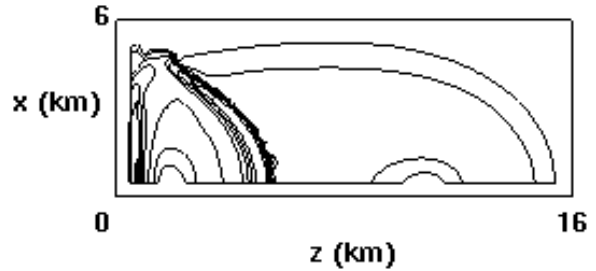}
\caption{ Contour lines of the log of the gradient of the rest mass
density $\log\left(\sqrt{\nabla^i(\rho)\nabla_i(\rho)}\right)$,
showing a shock front at coordinate radius $\sim 4.5 \; km$.  }
\label{fig:shock}
\end{figure}

In Fig.3 we show the position of the apparent horizon (AH).  To
confirm the location of the AH, convergence tests both in terms of
resolution and in terms of location of the computational boundary have
been carried out. (For a discussion of the AH finder, see
\cite{Alcubierre98b}).  We have also explicitly determined trapped
surfaces bounded by the AH for the positive confirmation of a
collapsed region.  In Fig. 3, the solid and long dashed lines
correspond to the AH locations at resolutions of $\Delta x = 0.492 \; km$
and $\Delta x = 0.246 \; km$,
while the dotted line corresponds to $\Delta x = 0.492 \; km$ but
with the outer boundary two times further out.  Although the coordinate
position of the AH is substantially elongated in the z direction, the
AH is actually quite spherical.  The proper
circumference on the x-y plane (equatorial) is close to
the circumference on the x-z plane (polar), with the latter being
$52.9 \pm 1.9 \; km$.  
For comparison, $4 \pi M_{AH}$ is $52.9 \pm 2.1 \; km$, where $M_{AH}$
is the mass of the AH (we note that a substantial part
of the matter in the system is enclosed within the AH).  
Analysis of this in relation to the hoop conjecture will
be given elsewhere.

Fig.4 shows the contour lines in the $y=0$ plane 
of the log of the gradient of the rest mass density 
$\log\left(\sqrt{\nabla^i(\rho)\nabla_i(\rho)}\right)$
at time $t =0.31 \; ms$.
We see a sharp peak at a coordinate radius of
$\sim 4.5 \; km$.  The sharp change in rest mass density 
indicates a shock, stronger in
the infalling direction ($z$), while weaker near the equatorial plane.  The
shock is moderately relativistic with Lorentz factor of about $1.2$.  The shock
is well captured in this $192^3$ run with high 
resolution shock capturing (HRSC) GR-hydro
treatment.  Comparing to Fig. 3, we see that the shock front is inside the AH
in all direction at this time, although it is still moving outward in coordinate
location.

% Figure
\begin{figure}
\psfig{figure=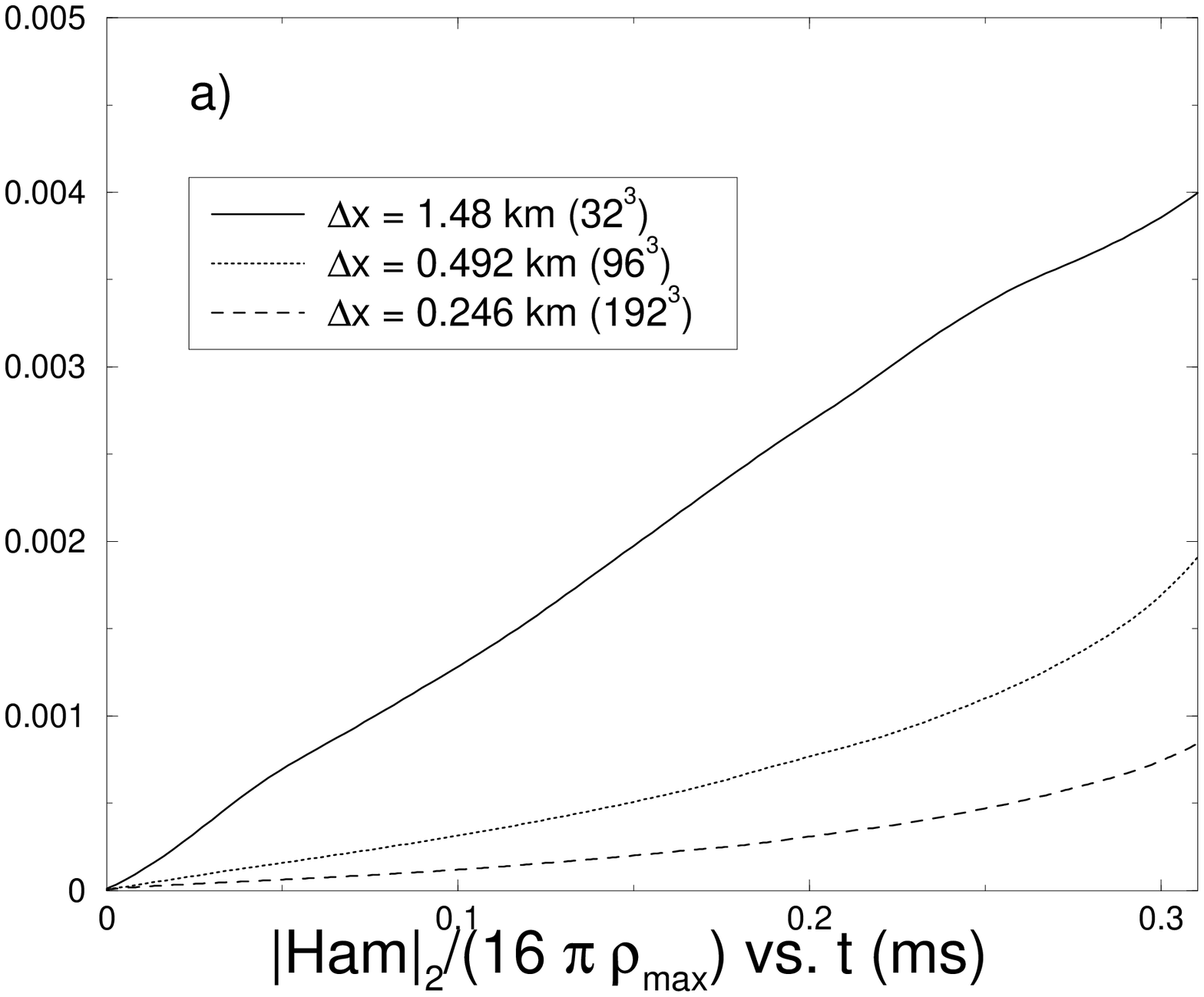,width=4.0cm}
%\vspace{-3.2cm}
%\hspace{-0.1cm}
\psfig{figure=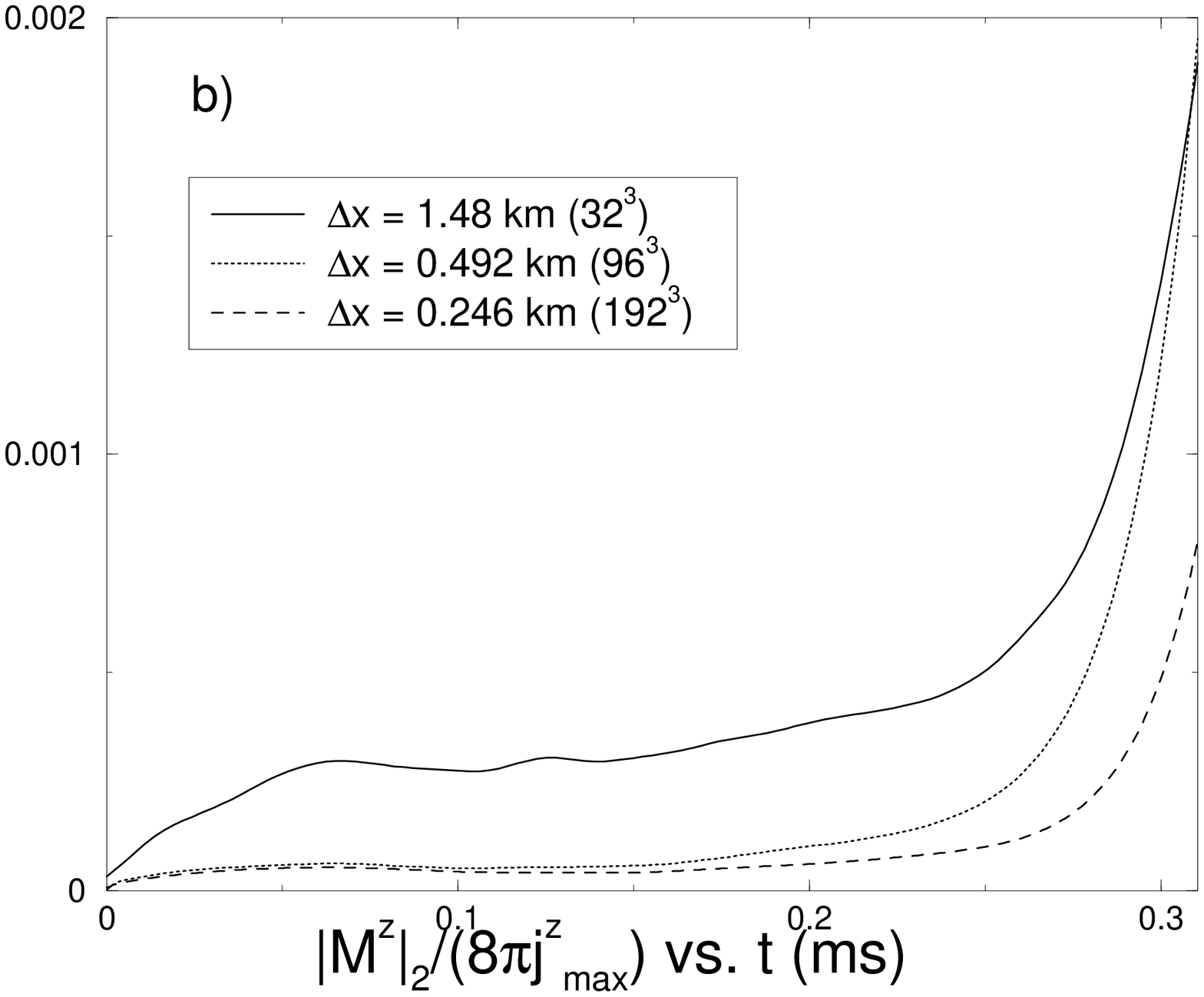,width=4.0cm}
\caption{ The evolutions of the $L2$ norms of (a) the Hamiltonian
constraint, and (b) the $z$-component of the momentum constraint.
}
\end{figure}

In Fig. 5 we show the convergence of the Hamiltonian and the
z-momentum constraints for a measure of the accuracy of the
simulation.  The evolution of the $L2$ norms (integrated
squared) of the constraints are scaled by the maximum of the matter
terms in the constraints ($16 \pi {\rho_{{}_{ADM}}}$ and $8 \pi j_{{}_{ADM}}^z$
respectively).  The solid, dotted, and dashed lines represent the
constraints at resolutions 
$\Delta x=1.48 \; km$, $0.492 \; km$ and $0.246 \; km$
respectively.  These {\it long time scale} convergence tests indicate
that our numerical evolution is {\it stable and convergent} for the
time scale of our present problem.  Towards the end we see that the
error is increasing rapidly; an examination of the spatial distribution of
the constraint violations shows that the error is due to the familiar
problem of resolving the ``grid stretching'' peaks of the black hole
metric (cf. Fig.2).  An extensive convergence analysis of many of the
variables involved in the simulation has been carried out and will be
presented in a follow up paper.  We have also performed simulations
with the initial boost velocity increased by $10\%$ (generating
more shock heating) and confirmed that our results are not sensitive
to the initial velocity.

With these results we conclude that prompt collapse of the merged object
formed in head-on collision infalling from infinity is possible, under
the {\it same} conditions as in Shapiro's conjecture.

We have also carried out simulations of head-on collisions of lower
mass NSs and have seen cases in which the shocks propagate
to cover the whole star and no AH is found, indicating that the collapse
would be delayed until radiative cooling.  A detailed analysis of the
transition point between prompt and delay collapse is computationally
expensive with our 3D code used to carry out the present analysis.  A
2D version of the present treatment is being developed with this
specific application in mind.

\paragraph*{\bf Conclusions.}
\label{conclusion}
We pointed out that there is an assumption in Shapiro's conjecture,
namely, the head-on collision process is in quasi-equilibrium (in the
sense of (A) and (B) above).  We showed that this may not be true for
the collision of two $1.4$ $M_{\odot}$ NS's.  We substantiated our
argument with a simulation solving the full set of the coupled
Einstein and general relativistic hydrodynamic equations.  We
confirmed the prompt formation of a black hole in the infalling time
scale $t_i$ with an apparent horizon found $0.16 \; ms$ after the
point of contact.

In this paper we concentrate on the head-on collision process under
the same conditions as in Shapiro's conjecture.  As the time scale
argument given above is rather general, and in particular does
not depend on the polytropc EOS, we expect the same argument to
be applicable to more general situations.  An investigation of the
prompt vs. dlayed collapse problem of head-on collisions with
realistic EOSs, more realistic initial conditions (initial data setup
with Post-Newtonian formulation), and with a determination of the
critical point between delayed vs. prompt collapse will be given in
follow up papers.

We thank all present and past members of our NCSA/Potsdam/Wash U team
for the joint ``Cactus'' code development effort, without which this
work would not be possible.  We thank in particular the 
contributions from Miguel Alcubierre for AH treatment and Bernd
Br\"ugmann for elliptic equation solvers used in this work.  We thank
Stu Shapiro for useful discussions.  This
research is supported by NASA NCS5-153, NSF NRAC MCA93S025, and
NSF grants PHY96-00507, 96-00049.

%%%%%%%%%bibliography%%%%%%%%%%%%%%
\bibliographystyle{prsty}
%\bibliography{references}

%\bibliography{bibtex/references}
\end{document}